\documentclass[aps,prd,twocolumn,showpacs,eqsecnum,nofootinbib]{revtex4}
\usepackage{amssymb}
\usepackage{amsmath}
\usepackage{fancybox}

\newcommand{\be}{\begin{equation} }
\newcommand{\ee}{\end{equation}}
\newcommand{\bea}{\begin{eqnarray}}
\newcommand{\eea}{\end{eqnarray}}

\newcommand{\ec}[1]{Eq.~(\ref{eq:#1})}

\newcommand{\eql}[1]{\label{eq:#1}}

\usepackage{graphicx}

\def\fun#1#2{\lower3.6pt\vbox{\baselineskip0pt\lineskip.9pt
  \ialign{$\mathsurround=0pt#1\hfil##\hfil$\crcr#2\crcr\sim\crcr}}}

\font\FermiSmallfont=cmssq8 scaled 1200

\def\LANLppthead#1#2#3{
\null 
\begin{center}\vskip -1.0truein{\hbox to 7.5truein {
\hfill
\vbox to 1in {\vfill \FermiSmallfont
              \hbox{#1}
              \hbox{#2}
              \hbox{#3}
              \vfill}
}}\vskip-0.0truein\end{center}}

\begin{document}

\LANLppthead {LA-UR-04-7094}{FNAL-PUB-04-348-A}{astro-ph/0411552}


\title{The Nonlinear Cosmological Matter Power Spectrum with Massive Neutrinos
  I: 
\\
The Halo Model} 

\author{Kevork~Abazajian$^1$, Eric~R.~Switzer$^{2}$, Scott~Dodelson$^{3,4}$,
  Katrin~Heitmann$^5$ and Salman~Habib$^1$}

\affiliation{$^1$MS B285, Theoretical Division, The University of
California, Los Alamos National Laboratory, Los Alamos, NM~~87545}
\affiliation{$^2$Department of Physics, Princeton University, Princeton, NJ~~08544}
\affiliation{$^3$NASA/Fermilab Astrophysics Center Fermi National Accelerator 
Laboratory, Batavia, IL~~60510}
\affiliation{$^4$Department of Astronomy \& Astrophysics, The University
of Chicago, Chicago, IL~~60637} 
\affiliation{$^5$MS D466, ISR-1, ISR Division, The University of
California, Los Alamos National Laboratory, Los Alamos, NM~~87545}

\date{January 24, 2005}

\begin{abstract}

Measurements of the linear power spectrum of galaxies have placed tight
constraints on neutrino masses. We extend the framework of the halo model of
cosmological nonlinear matter clustering to include the effect of massive
neutrino infall into cold dark matter (CDM) halos.  The magnitude of the effect
of neutrino clustering for three degenerate mass neutrinos with
$m_{\nu_i}=0.9\rm\,eV$ is of order $\sim$1\%, within the potential sensitivity
of upcoming weak lensing surveys.  In order to use these measurements to
further constrain---or eventually detect---neutrino masses, accurate theoretical
predictions of the nonlinear power spectrum in the presence of massive
neutrinos will be needed, likely only possible through high-resolution multiple
particle (neutrino, CDM and baryon) simulations.

\end{abstract}



\date{\today}
\maketitle

\section{Introduction}

The detection of neutrino flavor oscillation in conjunction with cosmological
arguments has highly constrained neutrino mass eigenvalues.  Solar
neutrinos~\cite{Fukuda:2001nk,Ahmad:2002jz} and atmospheric
neutrinos~\cite{Fukuda:1998mi,Fukuda:2000np} oscillate from one flavor to
another.  The KamLAND reactor neutrino detector has found evidence for neutrino
oscillations consistent with the inferred solar neutrino oscillation
parameters~\cite{Eguchi:2002dm}.  The K2K long-baseline experiment has found
evidence for neutrino oscillations consistent with the atmospheric
results~\cite{Ahn:2002up}. While flavor oscillation experiments constrain only
the neutrino mass differences, cosmological arguments have the advantage of
constraining the total mass. The present cosmological upper limits are
competitive with terrestrial experiments and expected to improve substantially
with time.

Massive neutrinos influence the large scale structures of the universe in a
well-defined way~\cite{Bond:1983hb} because they do not cluster, thereby
reducing the amount of matter that can accrete into potential wells.  The
galaxy power spectrum has been measured on large
scales~\cite{Percival:2001hw,Tegmark:2003uf} leading to upper limits on the sum
of neutrino mass ranging from $0.7$ to $1.8$ eV
\cite{Elgaroy:2003yh,Spergel:2003cb,Tegmark:2003ud,Abazajian:2004tn} (depending
on assumptions and data sets).  Combining estimates of the linear matter power
spectrum from the Lyman-$\alpha$ forest in the Sloan Digital Sky Survey (SDSS)
with estimates of the bias of galaxies in the SDSS with galaxy-galaxy lensing,
a tight limit of $0.42\rm\ eV$ is inferred for the 95\% C.L. limit on the sum
of three degnerate mass neutrinos~\cite{Seljak04param}.  The approximation of a
linear spectrum that is valid for such large scale measurements is no longer
valid on smaller scales where matter is highly clustered. Therefore, all
studies (except for Ref.~\cite{Abazajian:2004tn}) use data on the largest
scales.

Using small scale data in galaxy surveys requires knowing the nonlinear
clustering and bias of galaxies relative to the dark matter.  As we shall show,
the precision of the small scale galaxy clustering data such as that from the
SDSS is not high enough to warrant the inclusion of the effect of neutrino
clustering in dark matter halos, though other systematic effects may be
important (see Ref.~\cite{Abazajian:2004tn}). However, when large weak lensing
surveys -- which measure the mass distribution directly -- become available, it
will be essential to make direct use of this information, even on the smallest
scales due to the expected precision of their results~\cite{Refregier:2003xe}.

The nonlinear power spectrum for the dominant clustering component, cold dark
matter (CDM) itself has been best estimated by high resolution simulations by
the Virgo Collaboration~\cite{SPJWFPTEC}, but their work quanitifies the
uncertainty in their functional fits of the nonlinear power to approximately
7\%.  The effects of early free-streaming of neutrinos in suppressing
the linear power spectrum can be incorporated into predicting the nonlinear
power spectrum such as that from the fits of Refs.~\cite{PD96,SPJWFPTEC}.
However, one cannot na\"ively expect this to characterize the full effects of
massive neutrinos in the nonlinear regime.

Vale \& White~\cite{Vale:2003ad} showed that the uncertainties
introduced by approximations in ray-tracing techniques and numerical
convergence of pure dark matter simulations may be sufficiently reduced with
expected computing resources.  Similar to the effects of neutrino infall into
dark matter halos probed here, White~\cite{White:2004kv} and Zhan \&
Knox~\cite{Zhan:2004wq} showed that the effects arising due to baryonic cooling
and heating in CDM halos can alter the nonlinear matter power spectrum to
significantly alter the observed weak lensing signal.  Therefore, in order to
effectively use the information gained in upcoming weak surveys, one has to
accurately determine the nonlinear matter power spectrum for a given
cosmological model, and how it is influenced by the presence of baryons as well
as massive neutrinos.

In this paper, we first describe an analytic Boltzmann solution of neutrino
infall into cold dark matter (CDM) halos to calculate the modification of
captured neutrinos on the halos in \S\ref{clustering}.  We then employ the halo
model to calculate matter clustering statistics including the effects of
neutrino clustering, as well as the modification of the weak lensing power
spectrum while including or ignoring this effect in \S\ref{halomodel}, and then
sum up our conclusions in \S\ref{conclusions}.  In a companion
paper~\cite{paperII}, we use multiparticle numerical simulations to quantify
the effects of massive neutrino collapse into CDM halos on the nonlinear matter
power spectrum.

\section{Clustering of massive neutrinos in CDM halos}
\label{clustering}

To begin, we solve an isolated problem: how neutrinos cluster in the presence
of a dark matter halo. As we will see in the next section, the neutrino
clustering around halos at late times leads to changes in the nonlinear power
spectrum. These changes are unique signatures of massive neutrinos and may
eventually be detected.  A pioneering work~\cite{Brandenberger:1987kf} treated
the clustering of massive neutrinos around a point-like seed with the Boltzmann
equation.  While this seed was taken to be a cosmic string, their technique was
extended to accretion of neutrinos onto a CDM halo~\cite{Singh:2002de}.  The
density profile of a CDM halo has been found on average to follow a universal
profile over a wide range of mass scales, which we take to be a
Navarro-Frenk-White (NFW) form~\cite{Navarro:1996iw,Navarro:1997he}. The
structure of the inner portion of the profile is not crucial to the neutrino
clustering studied here.  One assumption used in Ref.~\cite{Singh:2002de} that
simplifies the Boltzmann solution to this problem is that the NFW CDM profile
is not influenced by the accretion of neutrinos; neutrinos do not act back on
the CDM.  (Formally, the source term in the Boltzmann equation is a pure
NFW-type CDM distribution plus the neutrinos).  This approximation is
reasonable because the neutrino mass associated with a cluster is a small
fraction of the mass in the cluster (of order 1\%, see below) and more
diffusely distributed, so that changes to the CDM NFW halo are essentially
negligible.

In the notation of Ref.~\cite{Brandenberger:1987kf}, the
Boltzmann solution for the Fourier transform of the neutrino profile
in an evolving CDM halo is:
\begin{widetext}
\begin{equation}
\label{turok_boltzmann}
\tilde \rho_{\nu} (k,\vartheta_i,\vartheta_f ) = \frac{ 4 G m_{\nu}^2 
T_{\nu,0}^2 (1+z)^3}{ \pi 
ka^3(\vartheta_f)} \int_{\vartheta_i}^{\vartheta_f} d \vartheta' 
a^4(\vartheta')  ( \tilde \rho_{\rm CDM} 
(k,\vartheta',M_{\rm CDM} (\vartheta')) + 
\tilde \rho_{\nu}
(k,\vartheta_i,\vartheta')) I \biggl (\frac{k(\vartheta_f - \vartheta') T_{\nu,0}
}{ m_{\nu}} \biggr ),
\end{equation}
\end{widetext}
where $a$ is the scale factor, but we parameterize integrals by 
a lookback super-conformal time, $d \vartheta = d \eta /a = d t /
a^2$; $T_{\nu,0}$ is the current neutrino temperature, $1.95$K; 
CDM evolution is included in the term $M_{\rm CDM} (\vartheta')$; and
\begin{equation}
I(\beta) \equiv \int_0^\infty dx \frac{x \sin (\beta x) }{ e^x + 1}.
\eql{defi}
\end{equation} 
The lower and upper limits of the $\vartheta$ integration set the time when the
accretion starts, $\vartheta_i$, and when it stops, $\vartheta_f$. The final
time is chosen to be at the redshift that the power spectrum is needed.  The
initial time is chosen to be early enough such that an extremely high fraction
of the neutrinos have velocities that cannot be captured by CDM halos, and deep
enough to be beyond the galaxy distrubution of upcoming weak lensing surveys,
at $z=5$.  Our results are essentially unchanged by choosing an initial $z=3$.

Before writing down the solution for the neutrino profile, we need to add one
ingredient to the earlier work~\cite{Brandenberger:1987kf,Singh:2002de}.  We
will be especially interested in the features of the power spectrum and its
suppression around $k=0.5 h^{-1}\rm\, Mpc$, near the transition from the 1-halo
to 2-halo terms in the halo model (see below).  Around $k=0.5 h^{-1}\rm\, Mpc$,
halos at and above $10^{14} h^{-1} M_\odot$ dominate the power
spectrum~\cite{Uros}.  These massive halos have collapsed only recently and
were significantly growing during the accretion evolution history of the
neutrinos.  To include this, we need to insert a growth factor in the source
term of the Boltzmann equation.  The merging and growth of CDM clusters was
studied extensively in Ref.~\cite{Wechsler}, who found that they evolve as:
\begin{equation}
\label{wechsler_evo}
M_{\rm CDM}(z) = M_{\rm CDM, today} e^{-2 a_c z},
\end{equation}
where the free parameter that defines the growth ($a_c$) is given by the
phenomenological relation $M_*(a_c) = 0.018 M_{\rm CDM, today}$
\citep{Wechsler}, where $M_*$ is the characteristic nonlinear mass, defined
where the fluctuation scale $\mu(m,z) = 1$ (see below).  That is, $a_c$ is the
scale factor at which a halo of mass $0.018 M_{\rm CDM, today}$ has collapsed.
For very large $k$, on the other hand, very light clusters dominate the power
spectrum and these clusters have collapsed in the distant past, so change very
little during the period from $z=1$ until today, when most neutrino accretion
takes place~\cite{Singh:2002de}.  For increasingly large scales, the evolution
of the CDM is significant: at the level of $2\%, 6\%, 15\%, 26\%$ for
$10^{12},10^{13},10^{14},10^{15}$ solar mass halos, respectively.

As it stands, the equation for the neutrino density transform
(\ref{turok_boltzmann}) cannot be simply integrated because the integral
contains $\tilde \rho_\nu$.  This term in the integral equation represents the
clustered neutrinos acting back on themselves: as more neutrinos accrete onto a
cluster, the cluster mass increases, and with it the ability to accrete more
neutrinos.  Since neutrinos make up a small fraction of the mass in a cluster,
the effect of neutrinos pulling in more neutrinos can also be ignored as a
second-order effect.

For cosmologies with more massive neutrinos that can make up a significant part
of the cluster mass, the nonlinear effects of the neutrinos acting back on
themselves and on the NFW-type CDM cluster may become important.  Ringwald \&
Wong~\cite{Ringwald:2004np} found that the effect of this neutrino
gravitational feedback is significant and the linearized Boltzmann equation
approach underestimates neutrino infall by a small amount for less massive
halos, but upto a factor of several for $\sim\!10^{15}M_\odot$ halos, which are
however rare.  The magnitude of the effect as estimated here is therefore a
minimum estimate of neutrino infall.  The magnitude of the effect of infall
will be quantified  in detail beyond such 1-halo approaches in the multiple
particle simulations in Ref.~\cite{paperII}.

The inner $I(\beta)$ is sometimes approximated by letting the Fermi-Dirac
denominator become a Boltzmann exponential \cite{Brandenberger:1987kf}; then
the integral can be performed analytically.  But Ref.~\cite{Singh:2002de}
points out that this approximation is off by $\approx 20 \%$ for large $k$, so
we use the full Fermi-Dirac type distribution.  

\begin{figure}[t]
\begin{center}
\includegraphics[width=3.25in]{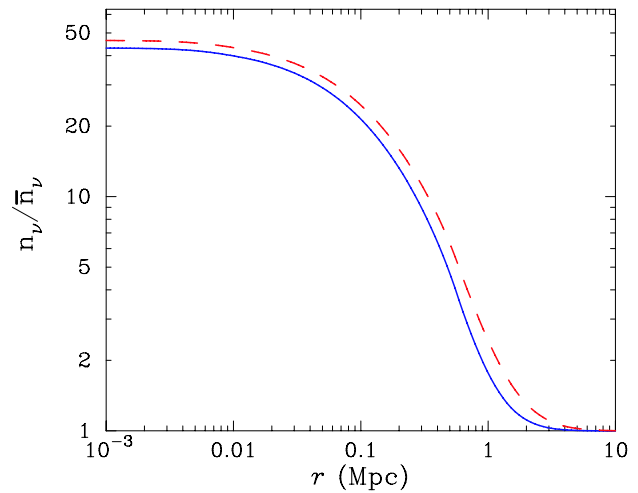}
\caption[]{\label{fig:evocomp}A comparison of the spatial distribution of accreted
  neutrinos for an NFW CDM profile for a $10^{14} h^{-1} M_\odot$ halo that
  has evolved in the fit of Ref.~\cite{Wechsler} (solid line) 
  and with a static profile (dashed line).  The virial radius for this halo is
  $4.8~\rm Mpc$ for a an $h=0.7, \Omega_c = 0.26$ cosmology.
}
\end{center}
\end{figure}

The neutrino profile around a $10^{14} h^{-1} M_\odot$ halo is shown in
Fig.~\ref{fig:evocomp}.  We see that neutrinos do not cluster in the halos as
efficiently as cold dark matter, so the profile inwards of $100$~kpc is mostly
flat and saturated (recall that the dark matter NFW profile increases as
$r^{-1}$ towards the center).  The CDM profile is truncated at one virial
radius.  The choice to cut off the CDM at its
virial radius suppresses the resultant neutrino population around $1\rm\ Mpc$
from the halo's center.

\section{The Halo Model}
\label{halomodel}

Here we show how to include the effects of massive neutrinos within the halo
model of nonlinear matter clustering and then estimate their magnitude.  The
halo model posits that all matter exists in halos and the correlations in the
matter can be explained by considering the correlations of the halos and the
density profile of matter within the halos. In the context of this model, there
are two places where neutrino masses affect the total mass distribution. First,
the linear power spectrum in a model with massive neutrinos differs from one
with massless neutrinos.  The linear power spectrum determines how halos are
correlated with each other. This first effect then is felt in the power
spectrum in the so-called {\it two-halo} term, the contribution of halo-halo
correlations to the total power spectrum.  This first effect is due to the
free-streaming of neutrinos on large scales, in the linear regime.  After CDM
halos form at redshift $z\lesssim 5$, massive neutrinos do not cluster as
efficiently as cold dark matter (again because of their velocities), but do
fall into the potential wells of CDM halos.  This second effect alters the
profile of the matter within a given halo.

\subsection{Overview}
Let us first review the halo model~\cite{peacock00,Uros,scoccimarro01}.
(See Ref.~\cite{Cooray:2002di} for a recent review.)  The nonlinear power spectrum
gets contributions from one- and two-halo terms: $P_{\rm NL}(k) = P_{1h}(k) +
P_{2h}(k).$; we first write them down and then explain the functions needed to
compute them:
\begin{equation}
P_{2h}(k) = P_{\rm lin}(k) \left( 
\int d \mu \frac{ f(\mu) b(\mu) 
\tilde \rho(k,\mu)}{ M(\mu) } \right)^2 ,
\label{2haloeq}\end{equation}
and
\begin{equation}
P_{1h}(k) = \int d \mu \frac{ f(\mu) }{\bar \rho M(\mu)} \left| \tilde
\rho(k,\mu)\right|^2 .\label{1haloeq}
\end{equation} 
Here $\bar \rho = \rho_{\rm crit} \Omega_c$ is the average matter density in
collapsed halos the universe, i.e. $\Omega_c$ is the fraction of the critical
density that is in halos, including neutrinos that have fallen into halos. Note
this is different from the usual halo model definition of $\bar \rho$ as the
total average mass density $\rho_{\rm crit} \Omega_m$, where neutrinos are
either not massive or ignored, and $\Omega_m = \Omega_{\rm CDM} + \Omega_b$.
Instead, we use $\Omega_c \equiv \Omega_{\rm CDM} + \Omega_b +\Omega_{\nu\rm
  halo}$. $\Omega_{\nu\rm halo}$ is calculated by integrating the mass of
neutrinos in individual halos over the halo mass function.

The halo model integrates over regions with overdensities parameterized
by\footnote{This ratio is usually denoted $\nu$, but we use $\mu$ here to save
  $\nu$ for neutrinos.} $\mu\equiv \delta_c^2/\sigma(M)^2$, with $\delta_c=1.68$
the linear overdensity at the epoch of a halo's collapse. Here we neglect the
effects of neutrino clustering on the definition of $\delta_c$.  
The rms of the linear fluctuations, $\sigma(M)$, filtered on a scale which on
average contains mass $M$ is
\begin{equation}
\sigma^2(M) = \int \frac{ d^3 k }{ (2 \pi)^3} P_{\rm lin} (k) | \tilde W_R (k)
|^2,
\label{sigma_squared}
\end{equation}
where $\tilde W_R (k)$ is the fourier transform of a top-hat window function of
size $R$, the radius enclosing mass $M$. Each mass $M$ then is associated with
a particular value of $\mu$.  Rare overdensities with large $\mu$
(high-$\sigma$ peaks) correspond to large mass halos. The correspondence can be
inverted to obtain $M(\mu)$, needed to compute the one and two- halo terms in
Eqs.~(\ref{2haloeq}) and (\ref{1haloeq}).

The number density of halos with mass $M$ is determined solely by the
dimensionless ratio $\mu$, which quantifies how rare the overdensity
is; we use the Sheth-Tormen distribution~\cite{Sheth:1999mn}
\begin{equation}
\frac{dn }{ dM} dM = \frac{ \bar \rho }{M} f(\mu) d\mu ,
\label{mass_nu}
\end{equation}
where
\begin{equation}
f(\mu) = A \frac{1 }{ \mu} (1+(\alpha \mu)^{-p}) \sqrt{\alpha
\mu} \exp (-\alpha \mu /2),
\label{sheth_thormen}
\end{equation}
with $A$ such that $f(\mu)$ integrated over all $\mu$ must be
equal to $1$ by mass conservation.  We adopt $\alpha =
0.707$ and $p = 0.3$.  The correlation between two halos depends
on their mass.  For cluster masses greater that $M(\mu=1)\equiv M_*$, halos
will be much more strongly clumped than surrounding matter.  This is
the nonlinear clustering regime, whose threshold is indirectly set by
the linear power spectrum.  The halo model bias, $b(\mu)$, folds in
this mass-dependent correlation into the overall correlation of two
halos, the "two halo" term.  The bias associated with the
Sheth-Thormen distribution is
\begin{equation}
b(\mu) = 1 + \frac{\mu -1 }{\delta_c} + \frac{2p }{ \delta_c (1+(\alpha \mu)^p)}.
\label{bias}
\end{equation}

\begin{figure}[t]
\begin{center}
\includegraphics[width=3.25in]{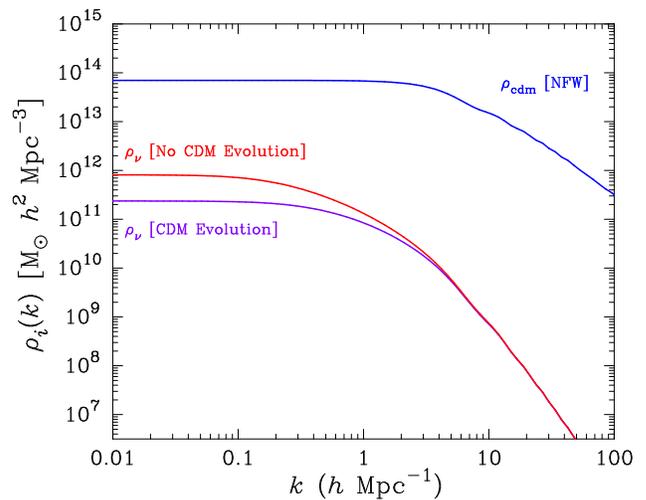}
\caption[]{\label{rho_nu_k}A comparison of the Fourier transform of the
  accreted neutrino profile into an NFW CDM halo of $10^{14} h^{-1}M_\odot$
  that has evolved with redshift (lower line) and with a static profile (upper
  line).  The Fourier transform for the CDM NFW profile is shown for reference
  of scale.  }
\end{center}
\end{figure}

\subsection{Modification for Neutrino Clustering}
\label{modification}
The components of the halo model described above can be readily updated if
neutrinos have mass. They depend solely on the linear power spectrum via
$\sigma$, and one can easily insert the effect of massive neutrinos on the
linear power spectrum. The final ingredient of the halo model is the mass
profile around halos, represented by its Fourier transform $\tilde\rho(k,\mu)$
in Eqs.~(\ref{2haloeq}) and (\ref{1haloeq}).  Usually, a natural choice is the
NFW form mentioned above,
\begin{equation}
\rho_{\rm CDM}(r,M) = \frac{\rho_s r_s^3}{r(r+r_s)^2}. 
\end{equation}
Here $r$ is the distance from the center of the halo (the conjugate variable to
$k$), $M$ is the halo mass, and the two parameters $\rho_s$ and $r_s$ are
functions of $M$ and the concentration $c$. In terms of these,
\begin{eqnarray}
\rho_s &=& \frac{ \Delta_v c^3(M) \bar \rho }{
3[\ln(1+c(M))-\frac{c(M) }{ 1+c(M)}]} \\
r_s^3 &=& \frac{ 3 M}{4 \pi c(M)^3 \Delta_v \bar \rho },
\end{eqnarray}
and $\Delta_v$ is the virial overdensity with respect to the mean matter density,
\begin{equation}
\Delta_v = \frac{18\pi^2+82x-39x^2}{1+x},
\end{equation}
and $x\equiv\Omega_c(z)-1$~\cite{Bryan:1998dn}.  We allow the concentration to
vary with redshift $z$ and halo mass \citep{Bullock:1999he}
\begin{equation}
c(M,z) = \frac{9}{1+z} \biggl( \frac{ M }{ M_* } \biggr )^{-0.13} ,
\end{equation}
where $M_*$ is the cosmology-dependent characteristic nonlinear mass scale.
Cluster evolution becomes nonlinear for virial masses greater than $M_*$.  The
halo model is simplest if we work in Fourier space, where a given spherically
symmetric density $\rho(r)$ becomes
\begin{equation}
 \tilde \rho(k) = \int_0^{r_{\rm cutoff}} (4 \pi r^2 dr) \rho (r) \frac{\sin (kr)
}{ kr} .\eql{trho}
\end{equation}
We choose this cutoff to be at the virial radius of a given cluster $r_{\rm
  vir} \equiv cr_s$.

While the NFW profile accurately describes the matter distribution if all
matter is cold, it does not account for infall clustering of neutrinos.  Since
neutrinos have non-zero thermal velocities even at latest times, they will not
strictly follow the cold dark matter profile in the halos. To account for this
aspect of neutrino infall clustering, we need to generalize Eqs.~\ref{2haloeq}
and \ref{1haloeq} by letting 
\begin{eqnarray} 
\tilde\rho(k,\mu) &\rightarrow&
  \tilde\rho_{\rm CDM}(k,\mu) + \tilde\rho_\nu(k,\nu) \\ M(\mu) &\rightarrow&
  M_{\rm CDM}(\mu) + M_\nu(\mu) .  
\end{eqnarray} 
Note that, for both cold matter and for neutrinos, 
\begin{equation} 
M_i = \frac{\lim_{k\rightarrow0} \tilde\rho_i(k)}{ (1+z)^3},
\end{equation} 
the factor of $1+z$ entering because $r$ in \ec{trho} is comoving distance.  

To include the effect of neutrino clustering, we add the clustered neutrino
mass profile $\rho_\nu(k,\mu)$ to the halo model's $k$-space density.  In the
spherical halo model, $\rho_\nu(k,\mu)$ can be approximated by solutions to the
Boltzmann equation for neutrinos around an NFW CDM halo of mass
$M(\mu)$. Fig.~\ref{rho_nu_k} shows a comparison of the Fourier transforms of
the NFW CDM halo profile and Boltzmann derived neutrino profiles.  The
$k\rightarrow 0$ limit of the Boltzmann Eq.~(\ref{turok_boltzmann}) gives
the total neutrino mass accreted onto a cluster, $M_\nu(\mu)$, which appears in
the halo model denominators, Eqs.~(\ref{final_1halo})-(\ref{final_2halo}).  We
do not include any potential changes the the form of the mass function
$f(\mu)$, since we assume the evolution of the CDM halos are unperturbed by the
presence of massive neutrinos.

\begin{figure}[t]
\begin{center}
\includegraphics[width=3.25in]{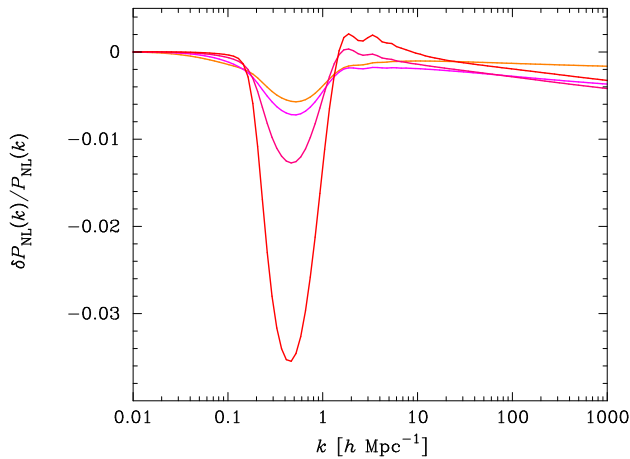}
\caption[]{\label{delta_pk} The estimated change in the nonlinear matter power
  spectrum at z=0 calculated with the halo model, $\delta P_{\rm NL}(k) =
  (P^\nu_{\rm NL}(k)-P^0_{\rm NL}(k))/P^0_{\rm NL}(k)$ due to late time
  neutrino clustering only.  (Both $P^0_{\rm NL}(k)$ and $P^\nu_{\rm NL}(k)$,
  are calculated with a transfer function that includes the same early neutrino
  free streaming.)  The lines of increasing magnitudes denote models with three
  neutrinos with masses of 0.1~eV, 0.3~eV, 0.6~eV, 0.9~eV, respectively.}
\end{center}
\end{figure}

With these modifications, the nonlinear power spectrum becomes
\begin{equation}
P_{\rm NL}^{{\rm CDM} + \nu }(k) = P_{1h}^{{\rm CDM} + \nu }(k) + P_{2h}^{{\rm
    CDM} + \nu }(k),
\end{equation}
\begin{widetext}
\begin{equation}
P_{1h}^{{\rm CDM} + \nu } (k) =  \int d \mu \frac{ f(\mu)}{\bar \rho
[M_{\rm CDM}(\mu)  + M_\nu(\mu)]} | \tilde \rho_{\rm CDM} (k,\mu) +  \tilde
\rho_{\nu} (k,\mu) |^2,
\label{final_1halo}
\end{equation}
and
\begin{equation}
P_{2h}^{{\rm CDM} + \nu} (k) = P_{lin}^{{\rm CDM} + \nu}(k) \left(\int{d \mu
\frac{f(\mu) b(\mu)}{M_{\rm CDM}(\mu) + M_\nu(\mu)} [\tilde \rho_{\rm CDM}
  (k,\mu) + \tilde \rho_{\nu} (k,\mu)]} \right)^2.
\label{final_2halo}
\end{equation}

\end{widetext}
We calculate the change in the nonlinear power spectrum due to these
modifications from neutrino clustering for a $\Lambda$CDM universe with
parameters $\Omega_{\rm cdm} = 0.26 -\Omega_\nu$, $h=0.7$, $\Omega_b = 0.04$,
$n=1$ and $\sigma_8=0.9$.  Massive neutrino models are chosen with three
degenerate mass neutrinos with $m_\nu=0.1, 0.3, 0.6, 0.9\rm\ eV$ (i.e., sum of
all neutrino masses of $0.3,0.9,1.8,2.7~\rm eV$), with $\Omega_\nu$ chosen
appropriately for these masses, while $\sigma_8$ is fixed.  The modification
$\delta P_{\rm NL}(k) = (P^\nu_{\rm NL}(k)-P^0_{\rm NL}(k))/P^0_{\rm NL}(k)$ is
shown in Fig.~\ref{delta_pk}, $P_{\rm NL}^0$ excludes the effects of late
neutrino clustering and $P_{\rm NL}^\nu$ includes them. Both $P^\nu_{\rm
  NL}(k)$ and $P^0_{\rm NL}(k)$ include the linear effects of early neutrino
free streaming, since we are interested in the bias imposed by ignoring the
effects of neutrino infall into CDM halos.  The drop in power at $k\sim 0.5
h\rm\ Mpc^{-1}$ occurs as expected at the scale where the most massive clusters
are contributing to the nonlinear power spectrum, and increases with increasing
neutrino mass.  The reduction in power is due to the smooth structure of the
accreted neutrino halo relative to the CDM halo.

\subsection{Weak Lensing Convergence Power Spectrum}

Planned weak lensing surveys have the potential to measure the power spectrum
very accurately.  Therefore, here we calculate the deviations due to massive
neutrino clustering on a weak lensing observable, namely the convergence power
spectrum, $C_{\ell}$.  This quantity is effectively the projected angular
matter-matter power spectrum weighted by the distribution of lensed galaxies.
The signal is estimated by~\cite{HuTegmark1999,Bartelmann:1999yn,Dodelson}
\begin{equation}
C_{\ell} = \frac{9}{16}\left(\frac{H_0}{c}\right)^4 \Omega_m^2 \int_0^{\chi_h}
{d\chi\ \left[\frac{g(\chi)}{a \chi}\right]^2 P\left(\frac{\ell}{\chi},z\right)},
\label{cl}
\end{equation}
for a universe with flat geometry, where $\chi$ is the comoving radial
distance, $\chi_h$ is the distance to the horizon, $a\equiv 1/(1+z)$, and
$P(k,z)$ is the nonlinear power spectrum at the appropriate redshift.  The weak
lensing weighting function is
\begin{equation}
g(\chi) = \chi \int_\chi^\infty {d\chi^\prime\ n(\chi^\prime)
  \frac{\chi^\prime-\chi}{\chi^\prime}},
\end{equation}
where $n(\chi)$ is the redshift distribution of the lensed galaxies normalized
such that $\int dz\, n(z)=1$.

The expected error in the observed weak lensing convergence power spectrum
comes from two sources: sample variance on large scales due to finite sky
coverage, and on small scales by the finite number density of galaxies on the
sky,
\begin{equation}
\Delta C_\ell = \sqrt{\frac{2}{(2\ell +1)f_{\rm sky}}} \left(C_\ell +
\frac{\gamma_{\rm rms}^2}{\bar n_{\rm gal}}\right),
\label{cl_error}
\end{equation}
where the fraction of sky covered $f_{\rm sky}$, intrinsic inferred galaxy
ellipticity $\gamma_{\rm rms}$, and average number density of galaxies $\bar
n_{\rm gal}$ are survey dependent.  

\begin{figure}[t]
\begin{center}
\includegraphics[width=3.25in]{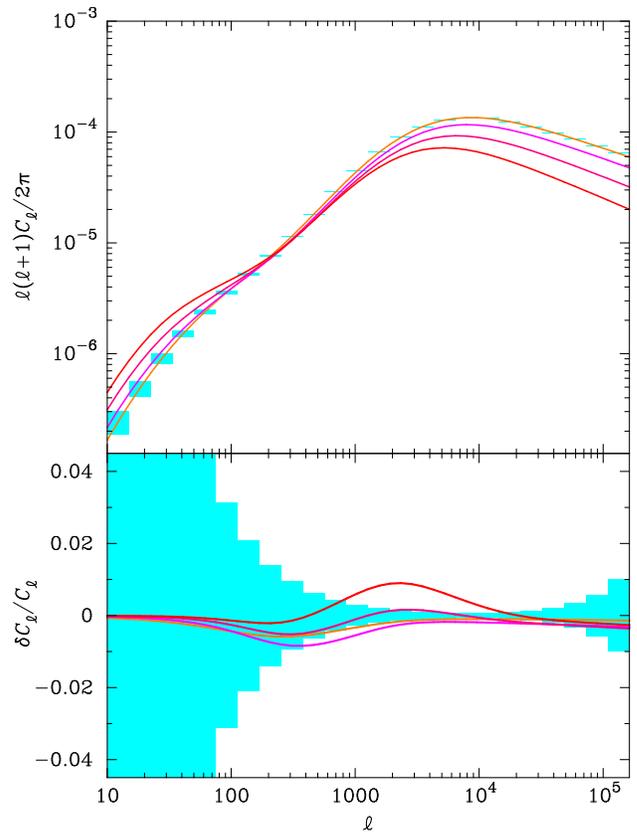}
\caption[]{\label{weak_cl_fig} The weak lensing convergence power spectrum
  (upper panel) for nonzero neutrino mass models of $0.1\rm~eV$, $0.3\rm~eV$,
  $0.6~\rm eV$, and $0.9~\rm eV$, with decreasing peak convergence,
  respectively.  The power spectra are normalized at $\sigma_8= 0.9$, therefore
  showing a pivot at $\ell \sim 200$.  The deviations including and excluding
  this effect are plotted in the lower panel, with increasing mass neutrinos
  corresponding to an increased amplitude of the effect. Gray (cyan) boxes are
  expected errors for an LSST-like survey as described in the text. }
\end{center}
\end{figure}

As an example, we use survey parameters similar to those possible with the
Large Synoptic Survey Telescope (LSST).\footnote{http://www.lsst.org/}
Specifically, the galaxy redshift distribution is of the form $n(z) \propto z^2
e^{-(z/z_0)^2}$, with $z_0=1$, average galaxy density $\bar n_{\rm gal} =
50\rm\ arcmin^{-2}$, a sky coverage of $f_{\rm sky}=0.5$, and $\gamma_{\rm
  rms}=0.15$.  In Fig.~\ref{weak_cl_fig} we show the effect on the weak
convergence power spectrum $\delta C_\ell = (C_\ell^\nu-C_\ell^0)/C_\ell^0$,
when we include ($C_\ell^\nu$) and exclude ($C_\ell^0$) the effect of neutrino
clustering in CDM halos to see the dependence of the weak lensing signal.  Note
that the linear effect of early neutrino free streaming is included in both
$C_\ell^\nu$ and $C_\ell^0$, since we are interested in the bias imposed by
ignoring the effects of neutrino infall into CDM halos.  The cosmological
parameters are chosen as in \S\ref{modification}.  Here, we take the neutrino
infall effect evaluated near the peak of the weak lensing weighting function
$g(\chi)$ at $z=0.4$, causing a change in the shape of the weak convergence
power spectrum deviation relative to the $z=0$ nonlinear power spectrum above.
The change in the nonlinear power spectrum affects the weak convergence power
spectrum from $\sim$0.1\% for three $0.1\rm\ eV$ neutrinos to $\sim$1\% for
three $0.9\rm\ eV$ neutrinos.

\section{Conclusions}
\label{conclusions}

We have estimated the effect of massive neutrinos in a concordance $\Lambda$CDM
cosmology on the nonlinear power spectrum in the halo model.  The shape of the
nonlinear power spectrum is changed due to neutrino infall in CDM halos at a
level of $\sim 0.5\%$ for three $0.1\rm\ eV$ neutrinos to $\sim 3\%$ for three
$0.9\rm\ eV$ neutrinos, corresponding to a change in the expected shape of the
weak convergence power spectrum at a level of $\sim 0.1\%$ for three
$0.1\rm\ eV$ neutrinos to $\sim 1\%$ for three $0.9\rm\ eV$ neutrinos, being
reduced due to the weight of higher redshift structures in cosmic shear
measurements.

Using the linear information from weak lensing surveys we may be able to constrain
the neutrino mass so that the effect of neutrino clustering may be
neglected~\cite{Abazajian:2002ck}.  Even given the inferred current upper
limits on neutrino masses from the linear power spectrum ($\Sigma m_{\nu_i}
<0.42\rm\ eV$, 95\% C.L.)~\cite{Seljak04param}, the effect of neutrino
clustering may be negligible unless current neutrino mass upper limits from the
linear power spectrum are too stringent, or new degeneracies may emerge among
features in the primordial power spectrum (e.g., running of the primordial
spectral index~\cite{Seljak:2004sj} or a break in the
spectrum~\cite{Blanchard:2003du}) which allow for larger neutrino masses.  

For more general cases where massive neutrinos are present in fits to observed
weak lensing convergence power spectra, massive neutrinos will need to be
included in numerical simulation predictions of the weak lensing signal in
addition to phenomena arising from baryonic condensation and
heating~\cite{White:2004kv,Zhan:2004wq}, leading to the potential necessity of
high-resolution multiparticle (neutrino, CDM and baryon) hydrodynamic numerical
simulations.  Coupled with upcoming weak lensing surveys, these predictions
will be a power probes of the contents of the cosmological soup as well as the
process of cosmological structure formation.

\acknowledgments We are grateful to Lam Hui, Dragan Huterer, Anatoly Klypin,
Lloyd Knox, Chung-Pei Ma, Andreas Ringwald and Tony Tyson for useful
discussions and comments on a draft.  KA, SH, and KH are supported by Los
Alamos National Laboratory (under DOE contract W-7405-ENG-36).  SD is supported
by Fermilab (under DOE contract DE-AC02-76CH03000) and by NASA grant
NAG5-10842.

\bibliography{v11}
\bibliographystyle{h-physrev4}

\end{document}